\documentclass{JFM-FLM_Au}

\usepackage{siunitx}
\usepackage{lineno}
\hypersetup{
    citecolor=blue,      
}

\lefttitle{L.-C. Huang and Y. Li}
\righttitle{Marangoni instability of cylindrical drops}

\title{Marangoni instabilities of cylindrical drops in a vertical Hele-Shaw cell immersed in stratified liquids}
 
\author{Li-Chen Huang\aff{1}, Yanshen Li\aff{1,2}}

\affiliation{\aff{1}School of Engineering Science, University of Chinese Academy of Sciences, Beijing 101408, PR China
\aff{2}State Key Laboratory of Nonlinear Mechanics, Institute of Mechanics, Chinese Academy of Sciences, Beijing, 100190, PR China}

\corresau{Yanshen Li, \email{liyanshen@ucas.ac.cn}}

\begin{document}
\maketitle

\begin{abstract}
The Marangoni instability of cylindrical drops in vertical Hele-Shaw cells immersed in stably stratified liquids has been studied previously, yet the underlying mechanism has not been explored thoroughly. Here we study the onset of the Marangoni instability of such a system by experimentally explore the parameter space of the drop radius and concentration gradient. The concentration field is directly observed with laser interferometry. The flow is found to become unstable when advection is too strong for diffusion to maintain a stable concentration field. However, two different instability regimes are found depending on the drop radius. When the drop is small, the friction force caused by the two plates of the Hele-Shaw cell is small so that it does not change much the velocity field. Marangoni advection in such a regime can be very strong so that the entire periphery of the drop can become unstable. When the drop is large, the friction becomes so large that the Marangoni velocity plateaus and the boundary layer thickness is also reduced. The modified velocity and concentration fields lead to another instability regime, where only liquid close to the equator of the drop becomes unstable. A unifying scaling theory that includes both instability regimes is developed, which agrees well with the experimental results. Our findings may shed new light on the understandings of Marangoni flows in confined geometries.

\end{abstract}


\section{Introduction}
\label{sec:Introduction}
Marangoni instabilities are triggered by interfacial tension gradients due to temperature or concentration gradients. After the pioneering work by \citet{benard1901tourbillons}, it has been studied extensively due to its important applications in liquid extraction \citep{sternling1959interfacial, jain2011recent}, melting \citep{schwabe1978experiments, chun1979experiments}, solidification \citep{dedovets2018five, meijer2023thin, wang2024self, van2024deforming} and coating techniques \citep{pearson1958convection, demekhin2006suppressing, wakata2025thermal}, etc. Marangoni instabilities also trigger interesting phenomena like Marangoni bursting \citep{keiser2017marangoni} and self-propulsion of droplets \citep{maass2016swimming, morozov2019self}. For a detailed review on Marangoni instabilities, we refer to \citet{nepomnyashchy2012interfacial}. For a better understanding on the physicochemical hydrodynamics of multicomponent systems \citep{levich1962physicochemical}, we refer to \citet{lohse2020physicochemical}.

The temperature/concentration gradients also lead to density gradients, which leads to the direct competition between Marangoni convection and gravitational convection. Though the gravitational effects are generally considered to be small as compared to interfacial tension gradients, it has been found recently that gravitational effects can be important or even dominating in some cases, such as in evaporating droplets \citep{edwards2018density, li2019gravitational, Diddens2021Competing, rocha2025evaporating}. For more specific cases where the density gradients are stable, intriguing phenomena also emerge, such as the continuous bouncing of a drop in a stable stratification \citep{li2019bouncing}. It has been found that the bouncing of the drop was triggered by two types of oscillatory Marangoni instabilities depending on the viscosity of the drop \citep{li2021marangoni, li2022marangoni}. However, the oscillatory flow induced by the instability cannot be clearly resolved because it was immediately disturbed by the drop's motion itself \citep{meijer2023rising}, thus hindering a deeper understanding of the instability. To (partially) resolve this issue, the position of the drop can be fixed by squeezing it in a vertical Hele-Shaw cell (which is also immersed in a stable stratification). Though by doing so, the geometry will be changed from spherical to cylindrical and the underlying physics might be different, it is still tempting to do so because of the following (additional) reasons.

First, the use of a Hele-Shaw configuration allows one to directly access the flow field information such as the evolution of the concentration field, which helps to better understand the instability mechanisms. Second, Marangoni flow in confined geometries are important for bubble/droplet manipulation \citep{basu2008virtual, gallaire2014marangoni, farzeena2023innovations, wang2024submegahertz}, micro mixing \citep{hu2017fluid, cha2021enhanced} and drug delivery \citep{stetten2018surfactant}, among others. Marangoni instabilities in Hele-Shaw cells have also been investigated for flat interfaces \citep{eckert2004chemical, cruz2021marangoni}, curved interfaces \citep{mokbel2017influence}, and drops/bubbles \citep{zuev2006oscillation, viviani2008experimental, schwarzenberger2015relaxation, bratsun2018adaptive}. Being one of the most basic configurations, however, the Marangoni instability of a cylindrical drop in a vertical Hele-Shaw cell immersed in a stable stratification has not been investigated throughly and the underlying physics remains unexplored. 

In this paper, we study the Marangoni instability in such a system by systematically exploring the two-dimensional (2D) parameter space spanned by the drop radius and the concentration gradient (made by ethanol-water mixtures). The concentration field around the drop is directly visualized by laser interferometry. Typical concentration fields of stable and unstable situations are shown, which indicates two different instability mechanisms. A unifying scaling theory which predicts two different instability mechanisms is then developed. Though both instability regimes originate from the competition between advection and diffusion, the friction force caused by the two plates of the Hele-Shaw cell is found to increase with the drop radius and consequently change the flow and concentration field, thus leading to two different instability regimes. Finally, the two instability criteria predicted from the scaling theory are compared with the experimental results and find good agreement. 

\section{Experimental procedure and methods}
\label{sec:ExpSetup}

\begin{figure}
  \centering
  \includegraphics[width=1\linewidth]{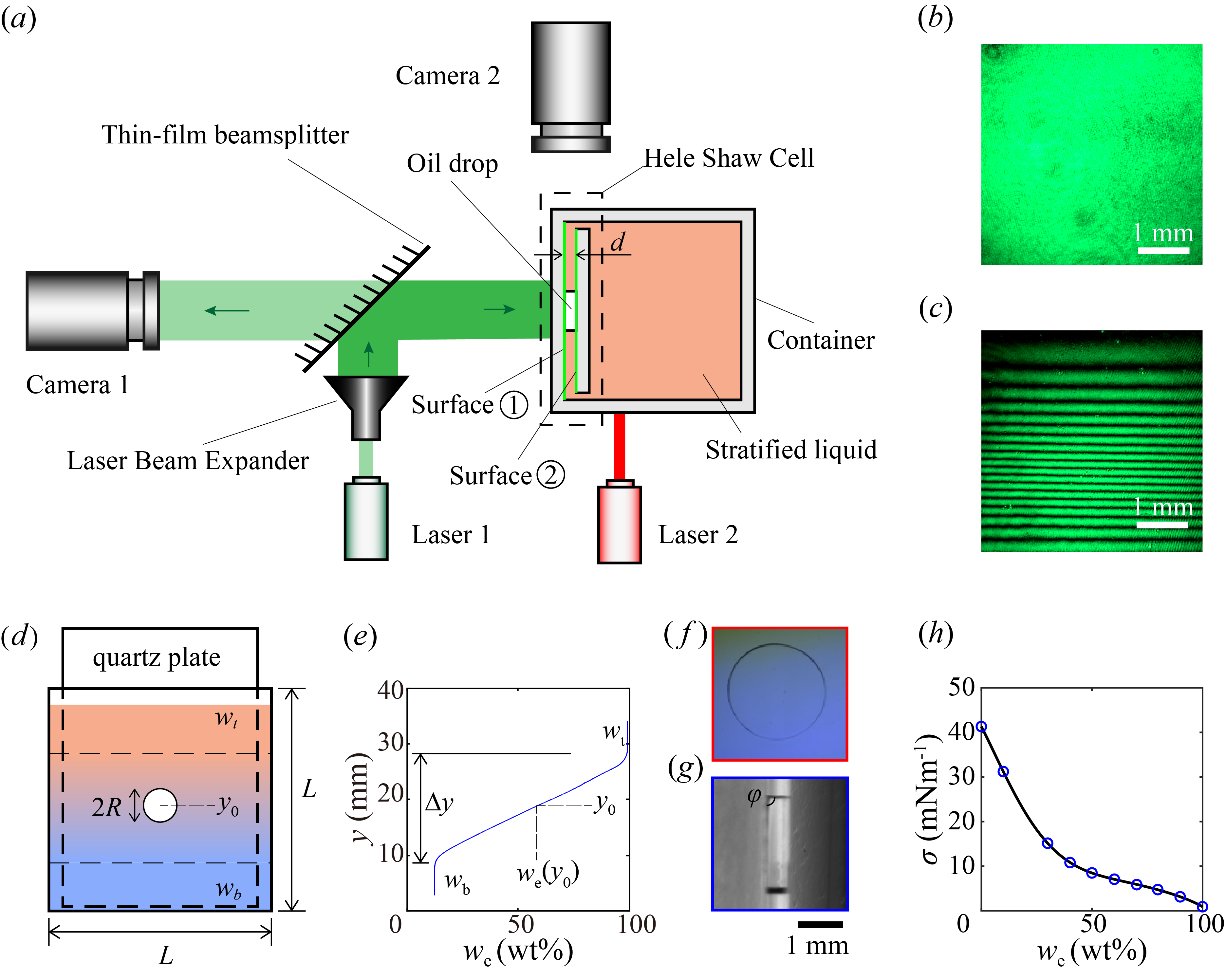}
  \caption{(\textit{a}) Sketch of the experimental set-up (top view). The Hele-Shaw cell was formed by a quartz plate and the front side of a cubic glass container. The interference fringes formed by the reflections from the two inner surfaces of the Hele-Shaw cell -- surfaces \protect\textcircled{1} and \protect\textcircled{2} -- were recorded by a front view camera. The thickness of the Hele-Shaw cell is $d$. A \SI{532}{nm} laser (Laser 1) was used to generated the interference patterns. The two surfaces \protect\textcircled{1} and \protect\textcircled{2} were adjusted parallel before it was filled with linearly stratified ethanol-water mixture. The exact ethanol concentration at different hight of the mixture was measured by laser deflection using Laser 2. A \SI{100}{cSt} silicone oil drop was injected in the Hele-Shaw cell and its side view was recoded by camera 2. ($b$) The interference pattern after the two surfaces \protect\textcircled{1} and \protect\textcircled{2} have been adjusted parallel but before the stratified liquid was injected. ($c$) The interference fringes when the Hele-Shaw cell was filled with linearly stratified liquids. Fringes indicate isopycnals of the stratified liquid. ($d$) A sketch of the front view of the oil drop and the Hele-Shaw cell filled with stratified liquid. The height of the center of the drop is $y_0$. ($e$) A typical measured ethanol concentration $w_\mathrm{e}$ as a function of height $y$. ($f, g$) Typical front \& side views of the oil drop. The contact angle of oil drop on surfaces \protect\textcircled{1} and \protect\textcircled{2} immersed in the stratified liquid is $\varphi$. The scale bar is \SI{1}{mm}. ($h$) Interfacial tension $\sigma$ of the \SI{100}{cSt} oil with the ethanol-water mixture at different ethanol concentrations.}
\label{fig:setup}
\end{figure}

To make the Hele-Shaw cell, a quartz plate was vertically held inside a cubic glass container (Hellma, 704.001-OG, Germany) close to its front side, see figure \ref{fig:setup}($a$) for a sketch of the experimental setup. The inner length of the cubic glass container is $L=\SI{40}{mm}$ and the size of the quartz plate is $50\times 35 \times \SI{1}{mm^3}$. Before injecting any liquid into the container and the Hele-Shaw cell, the two inner surfaces of the Hele-Shaw cell, surfaces \textcircled{1} and \textcircled{2}, were adjusted parallel using laser interferometry. A thick enough laser beam (of diameter $\approx\SI{10}{mm}$) was generated via a laser beam expander (Thorlabs, GBE05-A, USA) to cover as much area of the Hele-Shaw cell as possible. A relatively strong laser (Beijing Laserwave, LWGL532-100-SLM, China) of power \SI{100}{mW} was used to ensure the expanded laser beam was still intense enough to generate clear interference fringes. The expanded laser beam of wavelength \SI{532}{nm}  was guided to the Hele-Shaw cell via a thin-film beamsplitter (Thorlabs, BP245B1, USA). The coherent light reflected from surfaces \textcircled{1} and \textcircled{2} generated interference fringes, which were recorded by a camera (Nikon D850, \SI{30}{fps}, at $1920\times1080$ resolution) connected to a long working distance zoom lens system (Thorlabs, MVL12X12Z plus 0.25X lens attachment, USA). The quartz plate was adjusted until no fringes can be observed in the entire field of view (see figure \ref{fig:setup}($b$) for an example), then the two surfaces were considered to be parallel. The thickness of the Hele-Shaw cell was kept to be $d=0.5\pm 0.01 \,\si{mm}$ via a side view camera (Nikon D7500 at $1920\times1080$ resolution) connected to the same zoom lens system. No further adjustments to the Hele-Shaw cell were made in the subsequent experimental procedures.

Linearly stratified ethanol-water mixtures were injected in the glass container using the modified double bucket method, please see \citet{li2022marangoni} and \citet{meijer2023rising} for the detailed procedure. During the whole injection process, the glass container was covered by a lid to prevent the influence of preferential evaporation of the mixture and consequent flows. After the mixture has been injected, the Hele-Shaw cell was also filled with linearly stratified ethanol-water mixture. Now that surfaces \textcircled{1} and \textcircled{2} were parallel, the interference fringes indicated isopycnals, i.e., lines of equal ethanol concentration in the stratified mixture in between the two surfaces. Figure \ref{fig:setup}($c$) shows the typical isopycnals right after the stratified liquid has been injected. Within two minutes after the stratified liquid has been injected, the ethanol concentration $w_\mathrm{e}$ at different heights $y$ was measured by laser deflection \citep{lin2013one, li2019bouncing} using laser 2 (Ruichen, \SI{5}{mW}, China). A typical ethanol concentration profile is shown in figure \ref{fig:setup}($e$). Notice that two liquid layers of uniform concentration $w_\mathrm{t}$ and $w_\mathrm{b}$ were put on top and bottom of the linearly stratified liquid to facilitate the measurement of the ethanol concentration $w_\mathrm{e}$.

Then \SI{100}{cSt} silicone oil was carefully injected into the Hele-Shaw cell by a syringe (Hamilton, RN7636-01, USA) to form a cylindrical drop. The lid of the glass container was only removed during the injection of this oil drop. The drop radius $R$ was measured by taking a snapshot via the front view camera, see figure \ref{fig:setup}($d$) and ($f$) for the sketch and a typical snapshot. The height of the drop's center $y_0$ is controlled so that the ethanol concentration of the mixture at this height $w_\mathrm{e}(y_0)$ is kept within $\SI{40}{wt\%}\leq w_\mathrm{e}(y_0)\leq \SI{60}{wt\%}$. The contact angle $\varphi$ of the oil drop on surfaces \textcircled{1} and \textcircled{2} was monitored by taking a snapshot via the side view camera, see figure \ref{fig:setup}($g$). The two surfaces \textcircled{1} and \textcircled{2} of the Hele-Shaw cell has been chemically treated beforehand to control their hydrophobicity, so that $\varphi$ was mostly kept within $\SI{90}{}\pm \SI{5}{\degree}$, thus making the drop cylindrical, i.e., the oil-mixture interface of the drop is not protruding or concaving. Please see Appendix \ref{appA:detail} for the detailed chemical treating procedure and the statistics of $\varphi$. 

After the oil drop has been injected and its radius measured, laser 1 was turned on and the flow field information was recorded thereafter by the front view camera. It has been found that the ethanol concentration of the mixture does not change much in \SI{40}{min} \citep{li2022marangoni}, so each stratiﬁed mixture was used for no longer than 40 min. Otherwise, a new stratiﬁed liquid was prepared again. The time interval after the linear stratification has been generated till the end of the flow field recording was typically \SI{30}{min}, therefore, each stratified mixture was used for only one oil drop. For other drop radii $R$ and other concentration gradients $\mathrm{d}w_\mathrm{e}/\mathrm{d}y$, the aforementioned procedures of Hele-Shaw cell adjustment, stratified mixture generation, oil drop generation and subsequent recording were repeated. 

\section{Experimental results}
\label{sec:ExpResult}

\begin{figure}
  \centering\includegraphics[width=0.55\linewidth]{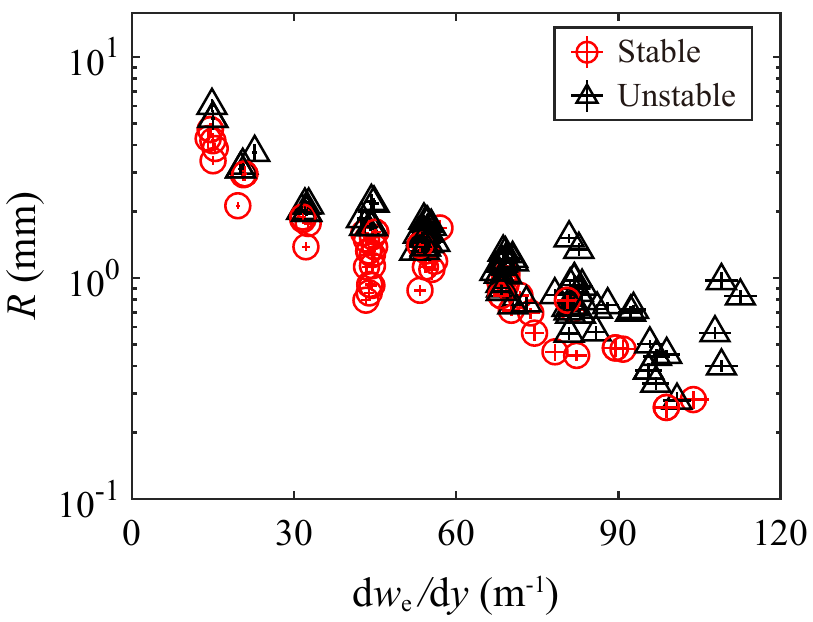}
  \caption{Phase diagram of the \SI{100}{cSt} drops with a drop radius $R$ versus concentration gradient $\mathrm{d}w_\mathrm{e}/\mathrm{d}y$ parameter space. Black triangles stand for unstable situations and red circles for stable situations.}
\label{fig:PhaseDiagram}
\end{figure}

As mentioned in Sec. \ref{sec:ExpSetup}, surfaces \textcircled{1} and \textcircled{2} has been adjusted parallel. Thus, when the ethanol concentration gradient is parallel to  surfaces \textcircled{1} or \textcircled{2}, the interference fringes indicate isopycnals, i.e, lines of equal density (or equal ethanol concentration). This holds true before the oil drop is injected. After injection of the drop, Marangoni advection induced by the drop might disturb the mixture so that it is possible that the ethanol concentration gradient is no longer parallel to surfaces \textcircled{1} and \textcircled{2}. However, since in most of the cases, the drop radius $R$ is (much) larger than the thickness of the Hele-Shaw cell $d$ (see figure \ref{fig:PhaseDiagram}), the concentration field could be considered as quasi 2D, thus we consider the fringes still represent the isopycnals when the oil drop is present. 

\begin{figure}
  \includegraphics[width=1\linewidth]{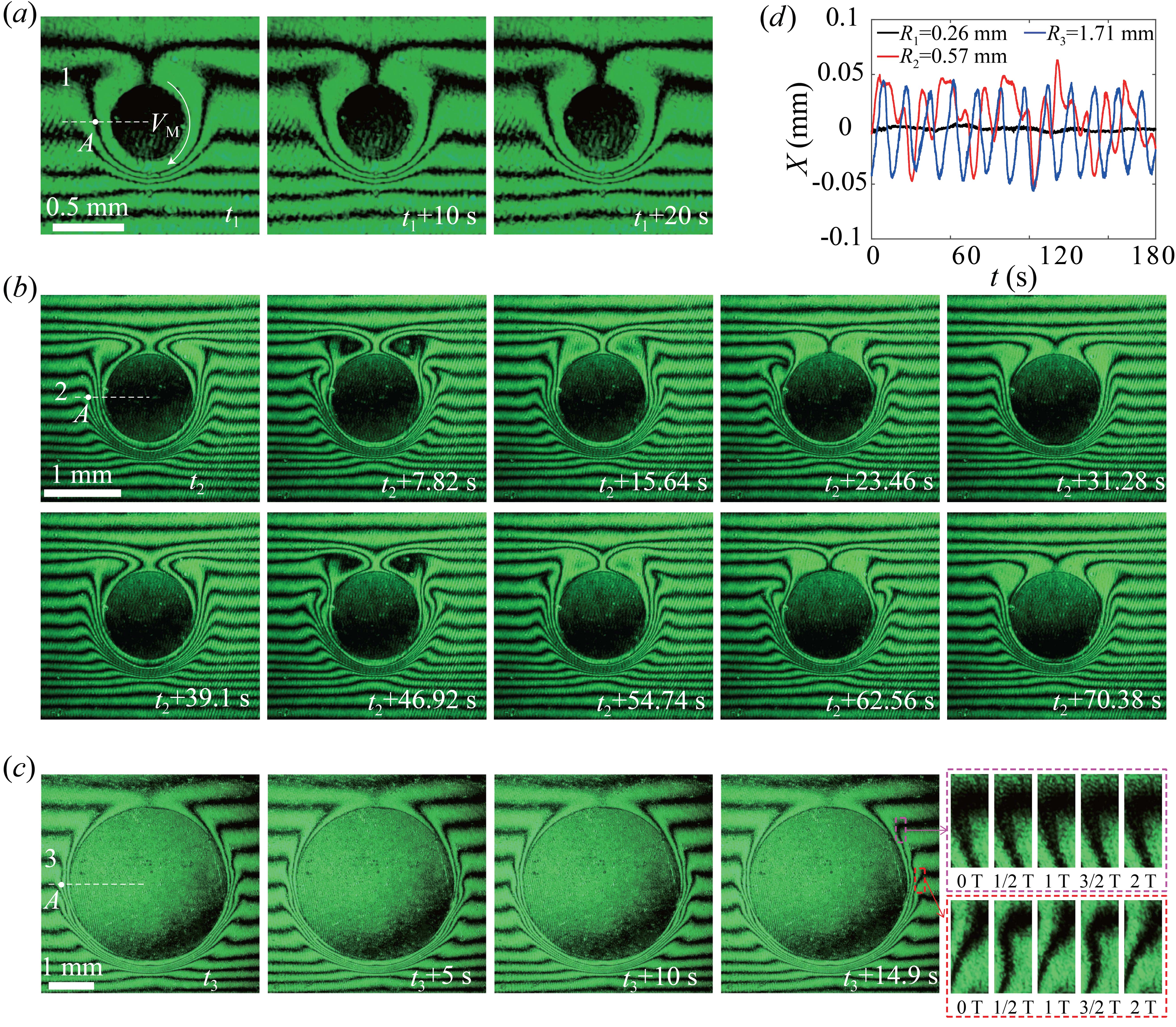}
  \caption{Snapshots of \SI{100}{cSt} silicone oil drops of different radii $R$ in linearly stratiﬁed ethanol–water mixtures with different ethanol concentration gradients $\mathrm{d}w_\mathrm{e}/\mathrm{d}y$. The snapshots start from some time $t_1$ (or $t_2$ or $t_3$) after the linear stratification has been generated. Point A is the intersection of a horizontal line passing through the drop's center and a dark fringe (labelled No. 1) right above this horizontal line. The isopycnals close to the drop all bend down because of the downwards Marangoni flow. (\textit{a}) $R=\SI{0.26}{mm}$, $\mathrm{d}w_\mathrm{e}/\mathrm{d}y=\SI{98.9}{m^{-1}}$, $t_1\approx\SI{1600}{s}$. The scale bar is \SI{0.5}{mm}. (\textit{b}) $R=\SI{0.57}{mm}$, $\mathrm{d}w_\mathrm{e}/\mathrm{d}y=\SI{107.9}{m^{-1}}$, $t_2\approx\SI{1500}{s}$. The oscillation period is $T=\SI{39.1}{s}$. The scale bar is \SI{1}{mm}. ($c$) $R = \SI{1.71}{mm}$, $\mathrm{d}w_\mathrm{e}/\mathrm{d}y=\SI{44.5}{m^{-1}}$, $t_3\approx1800\,\text{s}$. Zoomed views of the fringes close to the drop within an oscillation period $T=\SI{17}{s}$ are show in the dashed boxes. The fringes at the equator (red box) oscillates but the fringes above (magenta box) are stable. (\textit{d}) Temporal variation of the horizontal position $X$ of point A.}
\label{fig:FlowField}
\end{figure}

Typical fringe patterns for different drop radii $R$ in different concentration gradients $\mathrm{d}w_\mathrm{e}/\mathrm{d}y$ are shown in figure \ref{fig:FlowField}. For example, for a drop of radius $R=\SI{0.26}{mm}$ in a concentration gradient of $\mathrm{d}w_\mathrm{e}/\mathrm{d}y=\SI{98.9}{m^{-1}}$, the fringe patterns at different times are shown in figure \ref{fig:FlowField}($a$) (also see Supplementary Movie 1). Since the interfacial tension between silicone oil and ethanol is smaller than that of silicone oil and water (see figure \ref{fig:setup}($h$)), a downward Marangoni flow is generated on the surface of the drop, so that the isopycnals close to the drop are bended downwards. Far from the drop, the flow field is not disturbed, so that the isopycals are still horizontal, representing the background concentration gradient. Notice that over time, the isopycnals around this drop do not change, meaning that the flow field is stable. This is further confirmed by extracting the horizontal position $X$ of point A as a function of time (black line in figure \ref{fig:FlowField}($d$)), where point A is the intersection of a horizontal line passing through the drop's center and a dark fringe (labelled No. 1) right above this horizontal line, see figure \ref{fig:FlowField}($a$) for the definition of point A. 

When the drop is larger ($R=\SI{0.57}{mm}$) while keeping the concentration gradient almost unchanged ($\mathrm{d}w_\mathrm{e}/\mathrm{d}y=\SI{107.9}{m^{-1}}$), the isopycnals oscillate, indicating that the flow becomes unstable, see the snapshots in figure \ref{fig:FlowField}($b$) and also Supplementary Movie 2. Notice that in this case, the flow field around the entire periphery of the drop is unstable. The period of this typical oscillation is found to be $T=\SI{39.1}{s}$.

When the concentration gradient is smaller, the flow only becomes unstable for a larger drop, see figure \ref{fig:FlowField}($c$) and also Supplementary Movie 3 for an example where the concentration gradient is $\mathrm{d}w_\mathrm{e}/\mathrm{d}y=\SI{44.5}{m^{-1}}$ and the drop radius is $R=\SI{1.71}{mm}$. The period of this typical oscillation is $T=\SI{14.9}{s}$. Interestingly, contrary to figure \ref{fig:FlowField}($b$) where the entire periphery of the drop is unstable, here only the liquid close to the equator of the drop is oscillatory, see the zoomed snapshots in the right most panel. Liquid close to the drop but a bit higher than the equator is already stable. Naturally, liquid close to the top and bottom of the drop is also stable. 

The two different types of oscillations are also confirmed by the horizontal position $X$ of point A as a function of time, as seen in figure \ref{fig:FlowField}($d$). While in the first type the oscillation of $X$ is a bit complex (red line), in the second type its $X$ oscillates like a simple harmonic oscillator (blue line). The two types of oscillations indicate that there are two different instability mechanisms in this system.

\begin{figure}
  \centering\includegraphics[width=1\linewidth]{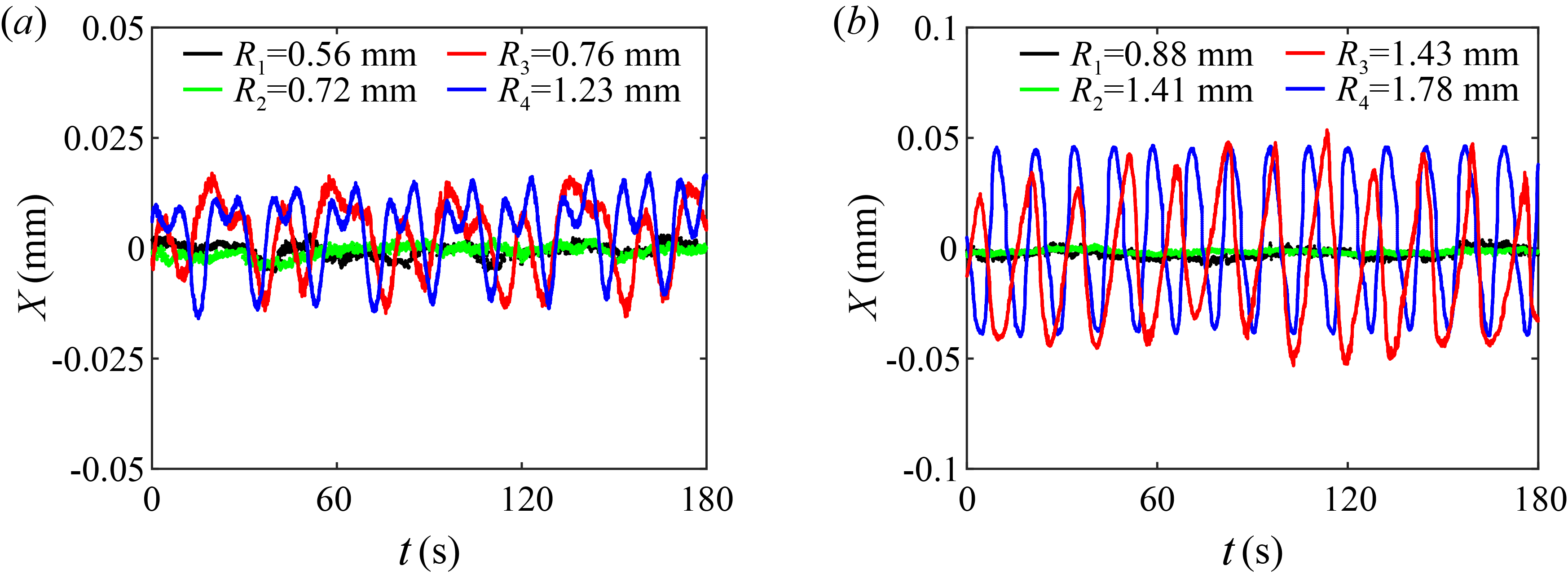}
  \caption{The variation of horizontal position $X$ with time $t$ for drops of different radii in similar concentration gradients: (\textit{a}) $\mathrm{d}w_\mathrm{e}/\mathrm{d}y\approx\SI{70}{m^{-1}}$, (\textit{b}) $\mathrm{d}w_\mathrm{e}/\mathrm{d}y\approx\SI{50}{m^{-1}}$.
   }
\label{fig:X}
\end{figure}

The horizontal position $X$ of point A as a function of time $t$ is further found to be a good indication of whether the flow is stable: As shown in figure \ref{fig:X}($a$) and ($b$), $X$ for drops of different radii in similar concentration gradients ($\mathrm{d}w_\mathrm{e}/\mathrm{d}y\approx\SI{70}{m^{-1}}$ and $\mathrm{d}w_\mathrm{e}/\mathrm{d}y\approx\SI{50}{m^{-1}}$) are plotted. It can be seen that only a small increase in the drop radius $R$ would lead to a dramatic increase in the oscillation amplitude of $X$. For example, for concentration gradient $\mathrm{d}w_\mathrm{e}/\mathrm{d}y\approx\SI{70}{m^{-1}}$, the oscillation amplitude of $X$ remains zero when the drop radius increases from \SI{0.56}{mm} to \SI{0.72}{mm}, but increases dramatically from zero to $\approx\SI{15}{\micro\meter}$ when the drop radius is further increased to \SI{0.76}{mm}. Similarly, for concentration gradient $\mathrm{d}w_\mathrm{e}/\mathrm{d}y\approx\SI{50}{m^{-1}}$, the oscillation amplitude of $X$ increases dramatically from zero to $\approx\SI{0.04}{mm}$ when the drop radius is increased from \SI{1.41}{mm} to \SI{1.43}{mm}. 

With this indicator, we further explore the onset of Marangoni instabilities of the drop in such a system by systematically varying the drop radius $R$ and the concentration gradient $\mathrm{d}w_\mathrm{e}/\mathrm{d}y$. The range of drop radius $R$ is varied between \SI{0.25}{mm} and \SI{6}{mm} and the range of concentration gradient $\mathrm{d}w_\mathrm{e}/\mathrm{d}y$ is varied between \SI{14}{m^{-1}} and \SI{113}{m^{-1}}. The results are shown in figure \ref{fig:PhaseDiagram}. It is found that the flow is easier to become unstable when the drop radius is larger or the concentration gradient is larger. In the next section, we will develop a unifying scaling theory to explain the two instability mechanisms.

\section{A unifying scaling theory for the onset of the instability}
\label{sec:Scaling theory for the onset of the instability}

\begin{figure}
  \centering\includegraphics[width=0.8\linewidth]{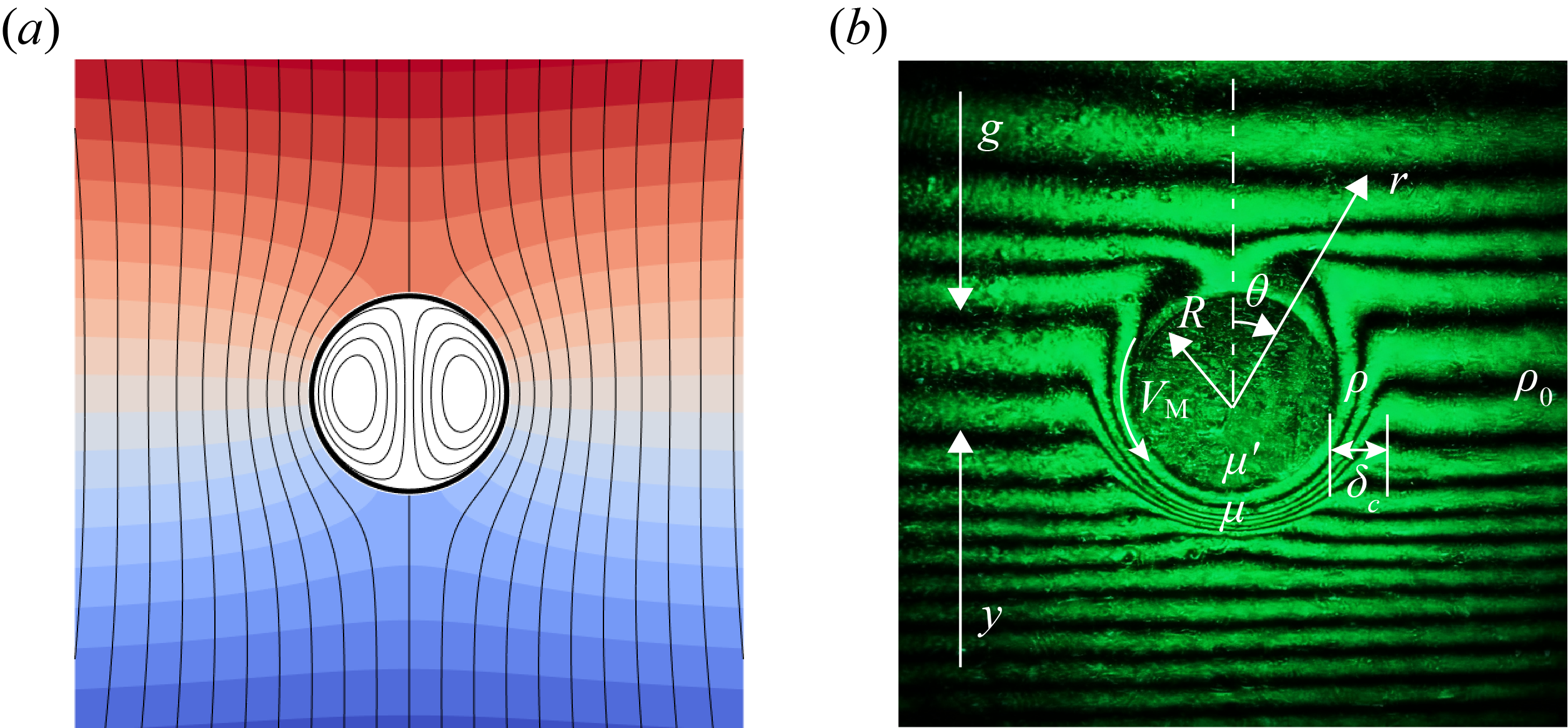}
  \caption{(\textit{a}) Streamlines and concentration field of a drop immersed in a stably stratified liquid in the ideal case, i.e., when $\mathrm{d}w_\mathrm{e}/\mathrm{d}y$ is very small so that the effect of gravity and advection (as compared to diffusion) are both negligible. Black lines represent streamlines and color strips represent isopycnals. (\textit{b}) Interference pattern for a drop of $R=\SI{0.26}{mm}$ at concentration gradient $\mathrm{d}w_\mathrm{e}/{\mathrm{d}y=\SI{98.9}{m^{-1}}}$. The flow is stable. Also shown is the definition of coordinates and physical properties. The plane polar coordinate $(r, \theta)$ has its origin at the center of the drop. The cartesian coordinate $y$ is pointing upwards and gravity $g$ is pointing downwards. The thickness of the concentration boundary layer is $\delta_c$.}
\label{fig:ConcentrationField}
\end{figure}

The system under investigation is a cylindrical drop in a Hele-Shaw cell vertically immersed in a linear stratification. Since the Hele-Shaw cell is very thin, the fluid velocity over the thickness of the cell has a Poiseuille profile. The thickness-averaged velocity $\mathbf{u}$ has analytical solutions in the ideal case, i.e., when the density gradient and advection (as compared to diffusion) are both negligible \citep{boos1997thermocapillary}. To display the analytical solutions, polar coordinate $(r,\theta)$ with the origin located at the center of the drop is used, see figure \ref{fig:ConcentrationField}($b$). The dimensionless stream functions inside $\tilde{\Psi}^\prime$ and outside the drop $\tilde{\Psi}$ are written as \citep{boos1997thermocapillary}

\begin{equation}
\tilde{\Psi}' = A_1 \left[ \frac{I_1(k\tilde{r})}{I_1(k)} - \tilde{r} \right] \sin(\theta),
\label{eq:PsiInside}
\end{equation}
\begin{equation}
\tilde{\Psi} = A_2 \left[ \frac{K_1(k\tilde{r})}{K_1(k)} - \frac{1}{\tilde{r}} \right] \sin(\theta),
\label{eq:PsiOutside}
\end{equation}
where $I_n$, $K_n$ are modified Bessel functions of the first and second kind of order $n$, respectively, $\tilde{r}=r/R$ is the dimensionless radial coordinate, $k=\sqrt{12}R/d$ is the dimensionless drop radius, $A_1$, $A_2$ are constants given by
\begin{equation}
A_1 = -I_1(k) K_0(k) / B,
\label{eq:A1}
\end{equation}
\begin{equation}
A_2= I_{2}(k)K_{1}(k)/B,
\label{eq:A2}
\end{equation}
and $B$ is given by
\begin{equation}
B = \tilde{\mu} (k^2 I_{1}(k) - 2k I_{2}(k)) K_{0}(k) + (1 - \tilde{\mu}) (k^2 K_{1}(k) + 2k K_{0}(k) I_{2}(k)),
\label{eq:C}
\end{equation}
where $\tilde{\mu} = \mu' / (\mu' + \mu)$, $\mu^\prime$ and  $\mu$ are viscosities inside and outside the drop. In the experiments, the viscosity of the drop ($\mu^\prime=\SI{96.6}{mPa\cdot s}$) is much larger than that of the mixture ($\mu<\SI{2.5}{mPa\cdot s}$), so that $\tilde{\mu}\approx1$. With this, the dimensionless tangential velocity $\tilde{u}_\theta$ outside the drop is obtained
\begin{equation}
\tilde{u}_\theta = -\frac{\partial \tilde{\Psi}}{\partial \tilde{r}} \approx \frac{I_{2}(k)K_{1}(k)}{[k^2 I_{1}(k) - 2k I_{2}(k)] K_{0}(k)}\left[ \frac{kK_{0}(k\tilde{r}) + \frac{1}{\tilde{r}}K_{1}(k \tilde{r})}{K_{1}(k)} - \frac{1}{\tilde{r}^{2}} \right] \sin(\theta).
\label{eq:DimensionlessUtheta}
\end{equation}
Then the dimensional tangential velocity is
\begin{equation}
u_\theta\approx-\frac{\mathrm{d}\sigma}{\mathrm{d}w_{\mathrm{e}}} \frac{\mathrm{d}w_{\mathrm{e}}}{\mathrm{d}y} \frac{R}{\mu + \mu'} \frac{I_{2}(k)K_{1}(k)}{[k^2 I_{1}(k) - 2k I_{2}(k)] K_{0}(k)}\left[ \frac{kK_{0}(k\tilde{r}) + \frac{1}{\tilde{r}}K_{1}(k \tilde{r})}{K_{1}(k)} - \frac{1}{\tilde{r}^{2}} \right] \sin(\theta).
\label{eq:Utheta}
\end{equation}
where $\mathrm{d}\sigma/\mathrm{d}w_{\mathrm{e}}$ is a material property (see figure \ref{fig:setup}($h$)) and $\mathrm{d}w_{\mathrm{e}}/\mathrm{d}y$ the undisturbed ethanol gradient in the far field. The Marangoni velocity at the equator of the drop is
\begin{equation}
\left. u_{\theta} \right|_{r = R, \theta = \SI{90}{\degree}} \approx-\frac{\mathrm{d}\sigma}{\mathrm{d}w_{\mathrm{e}}} \frac{\mathrm{d}w_{\mathrm{e}}}{\mathrm{d}y}  \frac{R}{\mu + \mu'} \frac{I_2(k)}{kI_1(k) - 2I_2(k)}.
\label{eq:Utheta90}
\end{equation}

As to the concentration field in the ideal case, advection is negligible as compared to diffusion, so that $\rho$ becomes a solution of Laplace's equation
\begin{equation}
\nabla^2\rho=0.
\label{eq:LaplaceEq}
\end{equation}
The linear density gradient in the far field is $\mathrm{d}\rho/{\mathrm{d}y}$. Denoting the far field density at the central height of the drop as $\rho_0$, the density field $\rho(r, \theta)$ can be written as
\begin{equation}
\rho(r, \theta) = \rho_0 + \frac{\mathrm{d}\rho}{\mathrm{d}y} r\cos \theta + f(r, \theta).
\label{eq:rhof}
\end{equation}
The second term on the right-hand side (RHS) ensures the density in the far field goes to the background density gradient $\mathrm{d}\rho/\mathrm{d}y$. The third term $f(r,\theta)$ describes the local density perturbation caused by the drop, which should approach $0$ when $r\rightarrow\infty$. Substituting (\ref{eq:rhof}) into (\ref{eq:LaplaceEq}), we have
\begin{equation}
f(r, \theta)=\frac{C}{r} \cos \theta,
\label{eq:f}
\end{equation}
where $C$ is a constant (with respect to $r$ and $\theta$) to be determined by the boundary condition ${\partial \rho}/{\partial r} = 0$ at $r = R$, which gives $C=R^2$. Substituting these into (\ref{eq:rhof}), we obtain
\begin{equation}
\rho = \rho_0 + \frac{\mathrm{d}\rho}{\mathrm{d}y} \left( r + \frac{R^2}{r} \right) \cos \theta.
\label{eq:rhoIdeal}
\end{equation}

The streamlines (\ref{eq:PsiInside}) and (\ref{eq:PsiOutside}) with $\tilde{\mu}=1$ and the density field (\ref{eq:rhoIdeal}) are shown in figure \ref{fig:ConcentrationField}($a$), where black lines represent streamlines and color strips represent isopycnals. 

However, when concentration gradient is not infinitely small, advection is not negligible and the density field deviates from the ideal case. As an example, figure \ref{fig:ConcentrationField}($b$) shows the interference pattern of a stable flow field when the concentration gradient is very large: $\mathrm{d}w_\mathrm{e}/{\mathrm{d}y=\SI{98.9}{m^{-1}}}$ and $R=\SI{0.26}{mm}$. It is found that isopycnals close to the drop are bent downwards because of the downward Marangoni flow. This is also the case for concentration gradients that are not so large, see another example shown in figure \ref{fig:FlowField}($c$) where $\mathrm{d}w_\mathrm{e}/{\mathrm{d}y=\SI{44.5}{m^{-1}}}$. As can be seen, the density field (\ref{eq:rhoIdeal}) loses its inversion symmetry around point ($\theta=\SI{90}{\degree}$, $\rho=\rho_0$). In this non-ideal case, the density field can be described by the density transport equation
\begin{equation}
\frac{\mathrm{D}\rho}{\mathrm{D}t} = D \nabla^2 \rho,
\label{eq:DensityTransport}
\end{equation}
where $D$ is the ethanol diffusivity. Expanding (\ref{eq:DensityTransport}) in the polar coordinate ($r$, $\theta$), we obtain
\begin{equation}
\frac{\partial \rho}{\partial t} + u_r \frac{\partial \rho}{\partial r} + \frac{u_\theta}{r} \frac{\partial \rho}{\partial \theta} = D \left( \frac{1}{r} \frac{\partial \rho}{\partial r} + \frac{\partial^2 \rho}{\partial r^2} + \frac{1}{r^2} \frac{\partial^2 \rho}{\partial \theta^2} \right).
\label{eq:DensityTransportPolar}
\end{equation}
For the liquid close to the equator of the drop, i.e., when $\theta=\SI{90}{\degree}$ and $r=R+\epsilon$ where $\epsilon$ is a very small length (for example, $\epsilon=\delta_c/2$ where $\delta_c$ is the concentration boundary layer, see figure \ref{fig:ConcentrationField}($b$)), the radial velocity $u_r$ is significantly smaller than the tangential velocity $u_\theta$. In the case under investigation, the flow is stable, so the density transport equation simplifies to:
\begin{equation}
\left.\frac{u_\theta}{r} \frac{\partial \rho}{\partial \theta}\right\vert_{r=R+\epsilon, \theta=\SI{90}{\degree}} \approx \left.D \left( \frac{1}{r} \frac{\partial \rho}{\partial r} + \frac{\partial^2 \rho}{\partial r^2} + \frac{1}{r^2} \frac{\partial^2 \rho}{\partial \theta^2} \right)\right\vert_{r=R+\epsilon, \theta=\SI{90}{\degree}}.
\label{eq:DensityTransportUtheta}
\end{equation}
Now that the Marangoni advection cannot be neglected and it tends to homogenize the concentration field around the drop (see figure \ref{fig:FlowField}), the concentration gradient close to the equator of the drop is smaller than the ideal case, which is $\frac{1}{r} \frac{\partial \rho}{\partial \theta}\vert_{r=R+\epsilon, \theta=\SI{90}{\degree}}\approx -2{\mathrm{d}\rho}/{\mathrm{d}y}$ from (\ref{eq:rhoIdeal}). Thus, in the non-ideal case, it is reasonable to assume that $\frac{1}{r} \frac{\partial \rho}{\partial \theta}\vert_{r=R+\epsilon, \theta=\SI{90}{\degree}}\sim{\mathrm{d}\rho}/{\mathrm{d}y}$. A smaller concentration gradient leads to a smaller Marangoni flow velocity as predicted by (\ref{eq:Utheta90}). Let $V_\mathrm{M}$ denote the Marangoni velocity close to the equator of the drop in our case, i.e., $V_\mathrm{M}=u_{\theta}|_{r = R+\epsilon, \theta = \SI{90}{\degree}}\approx u_{\theta}|_{r = R, \theta = \SI{90}{\degree}}$, it is reasonable to write
\begin{equation}
V_\mathrm{M}\sim-\frac{\mathrm{d}\sigma}{\mathrm{d}w_{\mathrm{e}}} \frac{\mathrm{d}w_{\mathrm{e}}}{\mathrm{d}y} R \frac{1}{\mu + \mu'} \frac{I_2(k)}{kI_1(k) - 2I_2(k)}.
\label{eq:VM}
\end{equation}
Note that $\frac{1}{r^2} \frac{\partial^2 \rho}{\partial \theta^2}\vert_{r=R+\epsilon, \theta=\SI{90}{\degree}}$ is negligible (see Supplementary Material for details). By replacing the left-hand side of (\ref{eq:DensityTransportUtheta}) and dropping the subscripts for convenience, we obtain 
\begin{equation}
V_\mathrm{M} \frac{\mathrm{d} \rho}{\mathrm{d}y} \sim D \left( \frac{1}{r} \frac{\partial \rho}{\partial r} + \frac{\partial^2 \rho}{\partial r^2} \right).
\label{eq:rhoReal}
\end{equation}

\begin{figure}
  \centering
  \includegraphics[width=1\linewidth]{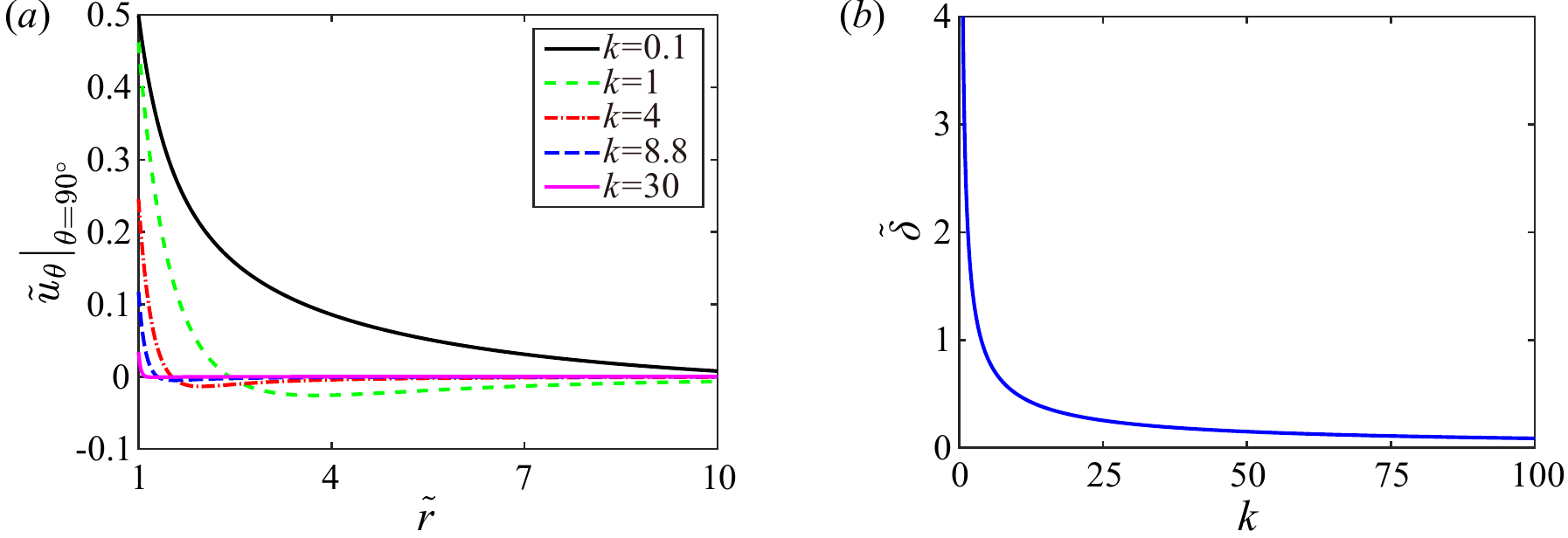}
  \caption{($a$) The dimensionless tangential velocity $\tilde{u}_\theta\vert_{\theta=\SI{90}{\degree}}$ (from (\ref{eq:DimensionlessUtheta})) as a function of $\tilde r$ at different $k$ values. The tangential velocity $\tilde{u}_\theta\vert_{\theta=\SI{90}{\degree}}$ first decreases, reaching a negative minimum value at $\tilde r_{u_{\theta, \mathrm{min}}}$ and then slowly approaches zero. The dimensionless tangential velocity at the equator of the drop $\tilde{u}_\theta\vert_{\tilde r=1,\theta=\SI{90}{\degree}}$ also decreases with $k$. ($b$) The dimensionless boundary layer thickness $\tilde\delta=\tilde r_{u_{\theta, \mathrm{min}}}-1$ as a function of $k$. Obviously, $\tilde\delta$ decreases with $k$.}
\label{fig:tildeUtheta}
\end{figure}

There are two terms in the RHS of (\ref{eq:rhoReal}) and we are interested in which one of them is larger. Since $\delta_c$ is the concentration boundary layer, we have $\partial\rho/\partial r\sim \Delta\rho/\delta_c$, where $\Delta\rho=\rho_0-\rho$ is the density difference between the liquid inside the boundary layer and in the far field. Also notice that (\ref{eq:rhoReal}) is evaluated close to the equator of the drop, we have $1/r\cdot\partial\rho/\partial r\sim 1/R\cdot\Delta\rho/\delta_c$. Similarly, we have $\partial^2\rho/\partial r^2\sim \Delta\rho/\delta_c^2$. Then the ratio of term one over term two becomes $\delta_c/R$. In our system, the density perturbation is mainly induced by Marangoni advection, see figure \ref{fig:ConcentrationField}, thus the concentration boundary layer thickness $\delta_c$ is proportional to that of the kinematic boundary layer thickness $\delta_v$ \citep{bejan1993heat}. Hence, we will not distinguish between them and use $\delta$ instead. Thus, the ratio of term one over term two is $\delta/R=\tilde{\delta}$. Then let us look at the characteristics of the flow field in the ideal case. From (\ref{eq:DimensionlessUtheta}), we plot $\tilde{u}_\theta\vert_{\theta=\SI{90}{\degree}}$ as a function of $\tilde r$ at different $k$ values, see figure \ref{fig:tildeUtheta}($a$). First, the tangential velocity at the equator of the drop $\tilde{u}_\theta\vert_{\tilde{r}=1, \theta=\SI{90}{\degree}}$ decreases with $k$. This is because $k$ actually reflects the friction force caused by the two plates of the Hele-Shaw cell \citep{boos1997thermocapillary}. The larger $k$, the more friction, thus the smaller $\tilde{u}_\theta\vert_{\tilde{r}=1, \theta=\SI{90}{\degree}}$. Second, for any given $k$, the tangential velocity first decreases with $\tilde r$ until it reaches a minimum value $\tilde{u}_{\theta,\mathrm{min}}\vert_{\theta=\SI{90}{\degree}}$ which is negative, later it increases slowly and approaches zero. The minimum value $\tilde{u}_{\theta,\mathrm{min}}\vert_{\theta=\SI{90}{\degree}}$ is negative because of mass conservation. It is also found that the radial position $\tilde r_{u_{\theta, \mathrm{min}}}$ of this minimum tangential velocity decreases with $k$. Given the velocity profile $\tilde{u}_\theta$, it is reasonable to define the (kinematic) boundary layer thickness as $\tilde\delta=\tilde r_{u_{\theta, \mathrm{min}}}-1$, where $\tilde\delta=\delta/R$. The boundary layer thickness $\tilde\delta$ is found to decrease with the dimensionless drop radius $k$, as shown in figure \ref{fig:tildeUtheta}($b$). This decrease is also caused by the increased friction at larger $k$. That is to say, the boundary layer is ``compressed'' because of the increased friction when the drop is larger.

\begin{figure}
  \centering\includegraphics[width=0.6\linewidth]{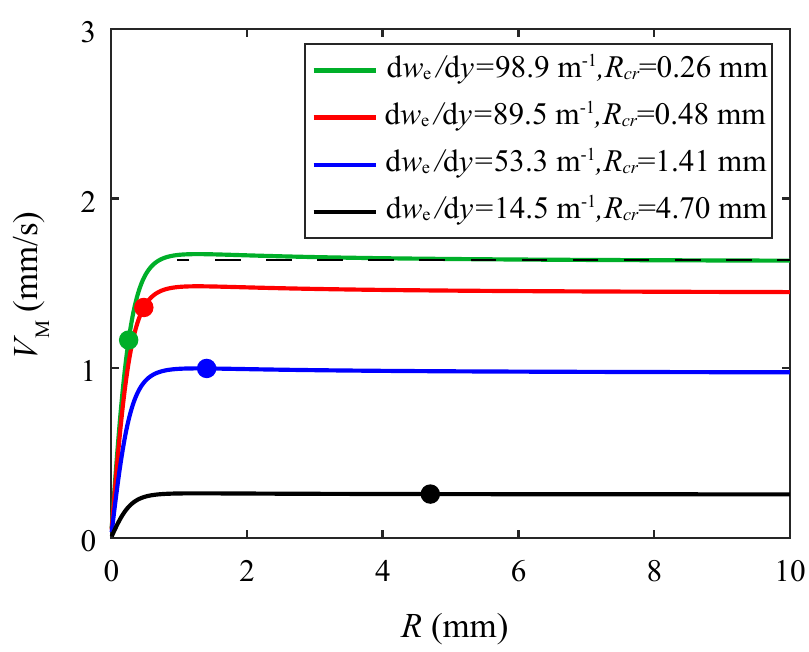}
  \caption{Marnagoni velocity $V_\mathrm{M}$ as a function of $R$ for different concentration gradients calculated from (\ref{eq:VM}). $V_\mathrm{M}$ first increases with $R$ until it reaches a maximum value at $R_\mathrm{m}=\SI{1.276}{mm}$, later it slowly decreases and approaches a constant value. The horizontal dashed line is a guide to the eye. Circular dots indicate the critical radius $R_\mathrm{cr}$ above which the flow becomes unstable.}
\label{fig:VM}
\end{figure}

Though this trend is found in the ideal case, it should follow in the non-ideal case, because in any case, the friction force would increase when the drop is larger. Consequently, we know that when the drop is small (large), the first (second) term on the RHS of (\ref{eq:rhoReal}) dominates. This is also confirmed by extracting the liquid densities close to the drop and comparing the two terms directly, see Supplementary Material for more details. From (\ref{eq:rhoReal}) we also know that as long as Marangoni advection $V_\mathrm{M}$ is strong enough, the flow will become unstable since diffusion cannot maintain a steady concentration field. We then check how $V_\mathrm{M}$ depends on $R$, see figure \ref{fig:VM} where $V_\mathrm{M}$ as a function of $R$ is plotted according to (\ref{eq:VM}). As can be seen, $V_\mathrm{M}$ first increases with $R$ until it reaches a maximum value at $R_\mathrm{m}=\SI{1.276}{mm}$ (or $k_\mathrm{m}=8.84$), later it slowly decreases and approaches a constant value. Apparently, the ``saturation'' of $V_\mathrm{M}$ at large $R$ is due to the increased friction. When the drop radius $R$ is small, friction is relatively small so that $V_\mathrm{M}$ can still increase with $R$. But when $R$ is large ($R>R_\mathrm{m}$), the friction is so large that $V_\mathrm{M}$ even starts to decrease. 
Since $V_\mathrm{M}\sim\mathrm{d}w_\mathrm{e}/\mathrm{d}y$, we know that for large concentration gradients, Marangoni advection $V_\mathrm{M}$ could be strong enough when the drop is small, thus the influence of friction is small and the first term in the RHS of (\ref{eq:rhoReal}) dominates. For small concentration gradients, the drop could easily become very large and the second term in the RHS of (\ref{eq:rhoReal}) dominates. The dots in figure \ref{fig:VM} denote the experimentally measured critical radius $R_\mathrm{cr}$ above which the flow becomes unstable at each concentration gradient. Indeed, the flow is unstable when the drop is small at large concentration gradients, and the flow only becomes unstable when the drop is very large at small concentration gradients. The two cases, i.e., at large or small concentration gradients, are considered subsequently in the following subsections.

\subsection{The limiting case when the concentration gradient is very large}
When the concentration gradient is very large, smaller drops are of interest so that $({1}/{r}) ({\partial \rho}/{\partial r})$ is dominant, then (\ref{eq:rhoReal}) reduces to
\begin{equation}
V_\mathrm{M} \frac{\mathrm{d} \rho}{\mathrm{d}y}\sim D  \frac{1}{r} \frac{\partial \rho}{\partial r}.
\label{eq:rhoRealLarge}
\end{equation}
From previous discussion we know that $1/r\cdot\partial\rho/\partial r\sim 1/R\cdot \Delta\rho/\delta$. Notice that the liquid close to the drop is brought down by the Marangoni flow from the top of the drop, we have $\Delta\rho\sim R\cdot\mathrm{d}\rho/\mathrm{d}y$, then $1/r\cdot\partial\rho/\partial r\sim 1/\delta\cdot \mathrm{d}\rho/\mathrm{d}y$. Substituting this into (\ref{eq:rhoRealLarge}), we have
\begin{equation}
V_\mathrm{M}\sim\frac{D}{\delta}.
\label{eq:rhoRealLarge1}
\end{equation}
Multiplying both ends of (\ref{eq:rhoRealLarge1}) with $R$ and rearrange, we obtain
\begin{equation}
\frac{V_\mathrm{M}R}{D}\sim\frac{R}{\delta}.
\label{eq:rhoRealLarge2}
\end{equation}
The left-hand side of (\ref{eq:rhoRealLarge2}) has the form of a P\'eclet number, which is referred to as the Marangoni number
\begin{equation}
Ma=\frac{V_\mathrm{M}R}{D}=-\frac{\mathrm{d}\sigma}{\mathrm{d}w_{\mathrm{e}}} \frac{\mathrm{d}w_{\mathrm{e}}}{\mathrm{d}y} R^2 \frac{1}{(\mu + \mu')D} \frac{I_2(k)}{kI_1(k) - 2I_2(k)},
\label{eq:Ma}
\end{equation}
where we have used (\ref{eq:VM}) with an equal sign. 
It has been known that $\delta \sim R/Ra^{1/4}$ for stable stratifications \citep{phillips1970flows, wunsch1970oceanic}, where
\begin{equation}
Ra = -\frac{\mathrm{d}\rho}{\mathrm{d}y} \frac{g R^{4}}{\mu D}
\label{eq:Ra}
\end{equation}
is the Rayleigh number for characteristic length $R$. Notice that now the friction is very small so that it does not change the scaling $\delta \sim R/Ra^{1/4}$. Substituting this and (\ref{eq:Ma}) into (\ref{eq:rhoRealLarge2}), also noticing that the flow becomes unstable when advection is too strong, the instability criterion thus is
\begin{equation}
{Ma}/{Ra^{1/4}} > s_1,
\label{eq:CriterionStrong}
\end{equation}
where $s_1$ is a constant to be determined.

\subsection{The limiting case when the concentration gradient is very small}
When the concentration gradient is very small, larger drops are of interest so that $\partial^2 \rho/\partial r^2$ dominates, then (\ref{eq:rhoReal}) reduces to
\begin{equation}
V_\mathrm{M} \frac{\mathrm{d} \rho}{\mathrm{d}y} \sim D \frac{\partial^2 \rho}{\partial r^2}.
\label{eq:rhoRealWeak}
\end{equation}
Now the friction is very large, the boundary layer is compressed so that its thickness does not follow $\delta\sim R/Ra^{1/4}$. However, since the concentration gradient is very small, the Marangoni velocity $V_\mathrm{M}$ is also very small so that the density field should be close to that of the ideal case, as shown by (\ref{eq:rhoIdeal}). Especially, the local density perturbation is more profound because the boundary layer is compressed. It is then reasonable to assume that the density profile close to the equator of the drop follows
\begin{equation}
\rho\sim\frac{\mathrm{d}\rho}{\mathrm{d}y}(\frac{R^2}{r}+ar),
\label{eq:SmallDensityProfile}
\end{equation}
where $a$ is a constant to be determined. This density profile is further confirmed by fitting with the density data extracted directly from the fringes, see Supplementary Material for more details. Substituting (\ref{eq:SmallDensityProfile}) into (\ref{eq:rhoRealWeak}), we obtain
\begin{equation}
V_\mathrm{M} \frac{\mathrm{d} \rho}{\mathrm{d}y} \sim D \frac{1}{R}\frac{\mathrm{d} \rho}{\mathrm{d}y}.
\label{eq:AdvDiffu}
\end{equation}
Reorganize, we have
\begin{equation}
Ma=\frac{V_\mathrm{M}R}{D}\sim s_2,
\label{eq:MaConst}
\end{equation}
where $s_2$ is a constant to be determined. When advection is too strong so that diffusion cannot maintain a stable concentration field, the flow becomes unstable. The instability criterion thus is
\begin{equation}
Ma>s_2.
\label{eq:CriterionWeak}
\end{equation}

In the next section, we will compare the two instability criteria mentioned above with the experimental results and measure the corresponding instability threshold $s_1$ or $s_2$. To calculate the Marangoni and Rayleigh numbers, ethanol weight fractions at the top of the drop $w_\mathrm{e}(y_0+R)$ are used to obtain the density $\rho$, viscosity $\mu$, diffusivity $D$ and the interfacial tension $\sigma$ (see \cite{li2022marangoni} for the concentration dependence of $\rho$, $\mu$ and $D$). 

\section{Comparison with experimental results and discussions}
\label{sec:Comparision}

\begin{figure}
  \centering\includegraphics[width=0.55\linewidth]{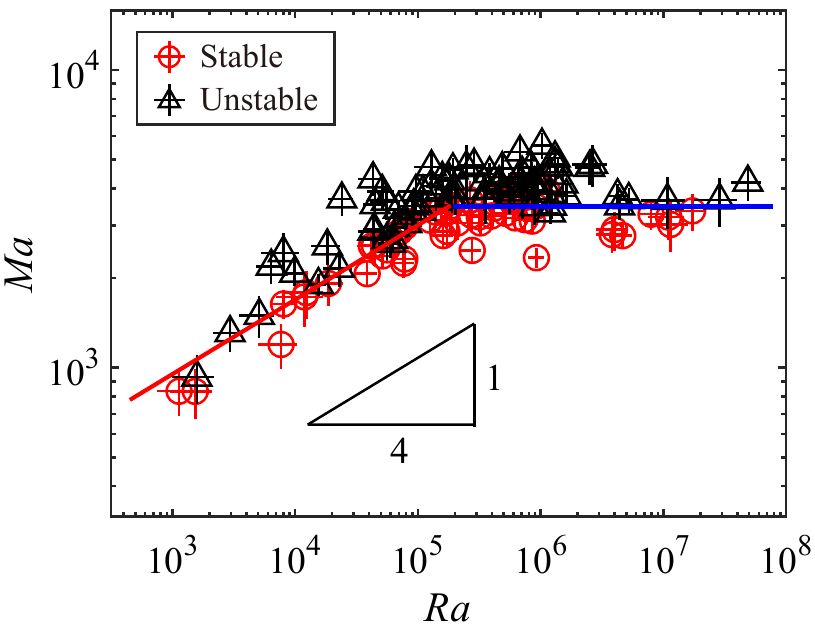}
  \caption{Phase diagram of the \SI{100}{cSt} drops replotted in the $Ma$ versus $Ra$ parameter space. Red circles stand for stable situations and black triangles for unstable situations.
The instability threshold starts from $Ma/Ra^{1/4} = 170$ for $Ra<\SI{1.89e5}{}$ (red solid line) and changes to $Ma=3490$ for $Ra>\SI{1.89e5}{}$ (blue solid line).}
\label{fig:MavsRa}
\end{figure}

The experimental results shown in figure \ref{fig:PhaseDiagram} are replotted in the $Ma$ versus $Ra$ parameter space in figure \ref{fig:MavsRa}. For large concentration gradients ($\mathrm{d}w_\mathrm{e}/\mathrm{d}y>\SI{70}{m^{-1}}$), the drops are small so that the Rayleigh number is small ($Ra<\SI{1.89e5}{}$), there is indeed a critical value $(Ma/Ra^{1/4})_\mathrm{cr}$ above which the flow becomes unstable (red solid line). The instability threshold in (\ref{eq:CriterionStrong}) is measured to be $s_1\approx170$. When the flow becomes unstable in this situation, liquid in the entire periphery of the drop is unstable, see figure \ref{fig:FlowField}($b$). This is because in this instability regime, the Marangoni advection is strong enough to induce large enough density perturbation to the entire periphery of the drop. If the drop's viscosity is smaller, Marangoni advection could become so strong that it would wrap up the drop in a liquid layer of almost uniform concentration, which is the case in Refs. \citet{zuev2006oscillation, viviani2008experimental, schwarzenberger2015relaxation, bratsun2018adaptive}. Thus, we call this the ``wrapping'' instability. For small concentration gradients ($\mathrm{d}w_\mathrm{e}/\mathrm{d}y\lesssim\SI{70}{m^{-1}}$), the drops are large so that the Rayleigh number is large ($Ra>\SI{1.89e5}{}$), there is indeed a critical Marangoni number $Ma_\mathrm{cr}$ (blue solid line) above which the flow becomes unstable. The instability threshold in (\ref{eq:CriterionWeak}) is measured to be $s_2\approx3490$. In this instability regime, the unstable region of the flow field is close to the equator of the drop and does not reach the drop's top and bottom. This is because the drop radius $R$ is large and the Marangoni velocity $V_\mathrm{M}$ has become ``saturated'' with $R$, thus advection cannot induce a large enough density perturbation across the entire drop periphery. Instead, the Marangoni advection is only strong enough near the equator of the drop, see (\ref{eq:DimensionlessUtheta}) and (\ref{eq:Utheta}). Thus, we call this instability the ``local'' instability.

It might seem surprising at first sight that such a simple system with only two independent input parameters ($\mathrm{d}w_\mathrm{e}/\mathrm{d}y$ and $R$) could have two different instability mechanisms. As a comparison, for the system of a spherical drop freely levitating/bouncing inside a linear stratification \citep{li2019bouncing, li2021marangoni, li2022marangoni}, there were three independent input parameters ($\mathrm{d}w_\mathrm{e}/\mathrm{d}y$, $R$ and drop viscosity $\mu^\prime$) and only two instability mechanisms were found. It turns out that the Hele-Shaw cell thickness $d$ serves as the third input parameter which was previously overlooked. The dimensionless drop radius $k=\sqrt{12}R/d$ also represents the friction force caused by the two plates of the Hele-Shaw cell. When $k$ is very large, the friction force modifies both the Marangoni velocity and the concentration field, eventually leads to another instability regime. 

\section{Conclusions and outlook}
\label{sec:Conclusion}
In summary, the Marangoni instabilities of cylindrical drops in a Hele-Shaw cell vertically immersed in a linearly stratified ethanol-water mixture were explored for different concentration gradients and drop radii. A unifying scaling theory was developed which predicts two different instability mechanisms, both of which originate from the competition between advection and diffusion. The friction caused by the two plates of the Hele-Shaw cell, represented by the dimensionless drop radius $k$, is the reason for the extra instability mechanism. (i) When the drop is small, friction is small so the Marangoni velocity increases with $R$. If the concentration gradient is large enough, advection can be strong enough to trigger instability. The instability criterion in this regime is $Ma/Ra^{1/4}>s_1$. (ii) When the drop is large, friction is so large that not only the Marangoni velocity $V_\mathrm{M}$ becomes ``saturated'', but also the boundary layer gets ``compressed''. This modified velocity and concentration field leads to another instability which is $Ma>s_2$. The scaling theory is well supported by the experimental results. In addition, in the first instability regime, the flow around the entire periphery becomes unstable when the criterion is met, this is because the concentration gradient is large so that the Marangoni advection can be very strong. We call this the ``wrapping'' instability. In the second instability regime, however, only the flow close to the equator of the drop becomes unstable when the criterion is met, this is because the Marangoni advection is relatively weak so that only advection near the equator can be strong enough to trigger the instability. We call this the ``local'' instability. 

An interesting feature of this system is the friction force due to the two plates of the Hele-Shaw cell, which increases with the dimensionless drop radius $k$. The increased friction leads to a decreased boundary layer thickness, it also makes the Marangoni velocity first increase with drop radius but then reach a plateau after reaching a peak velocity. The modified velocity field and concentration (if present) field due to the friction force should be noted in applications where the space is confined, for example, in droplet manipulation driven by Marangoni flows \citep{gallaire2014marangoni, luo2018marangoni}, or in Marangoni driven mixing in microfluidic devices \citep{bratsun2018adaptive, michelin2020spontaneous}.

\section*{Acknowledgements}
We thank Rama Govindarajan for valuable discussions. We acknowledge the financial support from the National Natural Science Foundation of China under grant No. 12272376.

\section*{Declaration of interests}
The authors report no conflict of interest.

\appendix
\section{Details of the surface treatment and the statistics of $\varphi$}
\label{app:detail}\phantomsection\label{appA:detail}
The two plates of the Hele-Shaw cell are both made of quartz (the cubic glass container is also made of quartz). If without surface treatment, the contact angle $\varphi$ (see figure \ref{fig:setup}($g$)) of silicone oil drops on the quartz surfaces immersed in ethanol-water mixtures would approach \SI{180}{\degree}, thus the silicone oil drops would adopt a ``pancake'' shape. To make the drops cylindrical, the contact angle $\varphi$ needs to be controlled around $\SI{90}{\degree}$ via surface treatment. We have tested several silanization procedures and found that methyltrichlorosilane works best \citep{wasserman1989structure}. The silanization procedure is described in the following.

The cubic glass container and the quartz plate are sequentially sonicated in acetone, isopropanol, ethanol, and deionized water for \SI{2}{min} each. After being dried with compressed air, they are treated by oxygen plasma (Harrick Plasma, PDC-002-HP, USA) at 30 watts for 2 hours, and immediately placed in a \SI{0.4}{\%} v/v solution of methyltrichlorosilane in octane for \SI{20}{min}. The samples are then immersed in chloroform for \SI{15}{min} to remove any residual organics or liquids. and in ethanol for \SI{15}{min}. They are finally dried with compressed air and ready to use.

\begin{figure}
  \centering\includegraphics[width=0.55\linewidth]{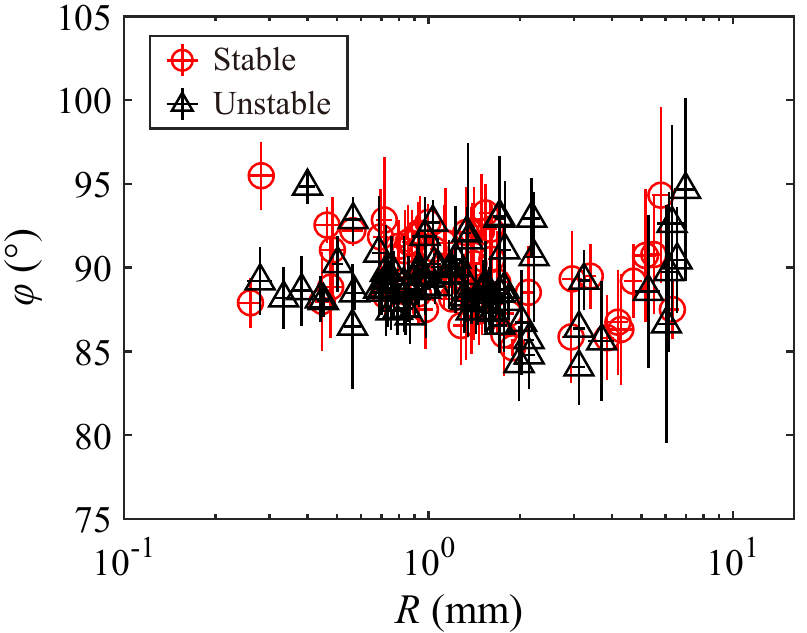}
  \caption{The contact angle $\varphi$ of the oil drop on surfaces \protect\textcircled{1} and \protect\textcircled{2} in stratified liquids for all the experiments. Circles and triangles represent the average value of $\varphi$ at the four corners, see figure \ref{fig:setup}($g$), and error bars represent the the standard deviation.}
\label{fig:Contactangle}
\end{figure}

Before each experiment, a side view of the silicone oil drop immersed in the stratified liquid is taken and the contact angle $\varphi$ of silicone oil on the chemically treated surface is measured by averaging the contact angles at the four corners. The contact angle $\varphi$ for each experiment is plotted versus the drop radius $R$, see figure \ref{fig:Contactangle}. It is found that $\varphi$ is controlled around $90\pm\SI{5}{\degree}$.

\bibliographystyle{jfm}
\bibliography{References}

\end{document}